\documentclass{article}
\usepackage{spconf,amsmath,graphicx}

\usepackage{multirow}
\usepackage{booktabs}
\usepackage{caption}

\usepackage{tikz}
\usetikzlibrary{positioning, shapes.geometric}

\title{PP-MeT: a Real-world Personalized Prompt based Meeting Transcription System}
\name{Xiang Lyu, Yuhang Cao, Qing Wang, Jingjing Yin, Yuguang Yang, Pengpeng Zou, Yanni Hu, Heng Lu}
\address{Ximalaya Inc., Shanghai, China}
\begin{document}
\copyrightnotice{979-8-3503-0689-7/23/\$31.00~\copyright2023 IEEE}
%\ninept
%
\maketitle
\begin{abstract}
Speaker-attributed automatic speech recognition (SA-ASR) improves the accuracy and applicability of multi-speaker ASR systems in real-world scenarios by assigning speaker labels to transcribed texts. However, SA-ASR poses unique challenges due to factors such as speaker overlap, speaker variability, background noise, and reverberation.
%This paper describes our PP-MET system in the Multi-channel Multi-party Meeting Transcription Challenge 2.0 (M2MeT2.0), which aims to solve the automatic speech recognition(ASR) task in a multi-party meeting scenario.
In this study, we propose PP-MeT system, a real-world personalized prompt based meeting transcription system, which consists of a clustering system, target-speaker voice activity detection (TS-VAD), and TS-ASR. Specifically, we utilize target-speaker embedding as a prompt in TS-VAD and TS-ASR modules in our proposed system. In constrast with previous system, we fully leverage pre-trained models for system initialization, thereby bestowing our approach with heightened generalizability and precision.
%Specifically, our proposed integrated PP-MET system consists of a clustering system, target-speaker voice activity detection (TS-VAD), and target-speaker ASR (TS-ASR). Moreover, we utilize personalized speaker embedding as a prompt in TS-VAD and TS-ASR in our proposed integrated meeting transcription system. To better learn the ..., the pre-trained Model is used to ...
Experiments on M2MeT2.0 Challenge dataset show that our system achieves a cp-CER of 11.27\% on the test set, ranking first in both fixed and open training conditions.
\end{abstract}
\section{Introduction}
\label{sec:intro}

%With the advances in deep learning, lots of progress has been achieved in automatic speech recognition (ASR) and its performance has been improved significantly.
%However, ASR systems are still susceptible to performance degradation in real-world far-filed scenarios like meetings or home parties, due to the background noise, inevitable reverberation, and multiple speakers overlapping.

The rapid advancements in deep learning have led to remarkable strides in automatic speech recognition (ASR), substantially enhancing its overall performance. Despite these achievements, ASR systems continue to face challenges in real-world far-field scenarios, such as meetings or home parties, where background noise, unavoidable reverberation, and overlapping speech from multiple speakers can significantly degrade their performance. 
%In response to these obstacles, researchers have devoted considerable attention to the development of robust ASR systems for such demanding acoustic environments. Specifically, numerous studies have concentrated on multi-microphone multi-party speech recognition and diarization, particularly in the context of dinner party scenarios. The goal is to devise efficient methods to accurately transcribe and distinguish speakers in these complex and challenging settings.
In order to develop a robust ASR system in such challenging acoustic environments, numerous research studies have concentrated on multi-channel multi-party speech recognition and diarization within dinner party scenarios~\cite{barker2018fifth, watanabe2020chime}.

%The M2MeT2.0 challenge~\cite{yu2022m2met} aims to solve the task of Automatic Speech Recognition(ASR) in a multi-party meeting scenario, necessitating the provision of accurate transcriptions and identification of the corresponding speakers.
The objective of the M2MeT2.0 challenge~\cite{yu2022m2met,liang2023second} is to address the ASR task in multi-party meetings, which involves providing precise transcriptions and identifying the corresponding speakers.
To advance the practical application of current multi-speaker speech recognition systems, the M2MET 2.0 Challenge evaluates the task of Speaker-attributed ASR (SA-ASR). Additionally, the challenge includes two sub-tracks: fixed training condition track and open training condition track.
Speaker-attributed ASR (SA-ASR) poses several challenges due to the complexity of accurately attributing speech to specific speakers. 
The SA-ASR task improves the accuracy and applicability of multi-speaker ASR systems in real-world scenarios by assigning speaker labels to transcribed texts. Unlike traditional ASR systems that transcribe speech without considering speaker identities, SA-ASR goes a step further by associating each recognized word or phrase with the corresponding speaker.

%The goal of SA-ASR is to provide not only accurate transcriptions of spoken content but also information about which speaker produced each segment of speech. This attribution of speech to specific speakers can be valuable in various applications, such as meeting transcription, speaker identification, speaker turn-taking analysis, or speaker-specific information extraction.

SA-ASR faces unique challenges due to factors like speaker overlap, speaker variability, background noise, and reverberation. Overcoming these challenges involves developing advanced algorithms and techniques for speaker diarization, speech separation, and speaker recognition to accurately attribute spoken words to their respective speakers.
The development of SA-ASR systems has the potential to improve the performance and usability of speech recognition in scenarios where multiple speakers are present, enabling applications that require speaker-specific information and analysis.

In this study, we present the PP-MeT system, a personalized-prompt based meeting transcription system designed to address the ASR task in multi-party meetings. Our approach comprises three essential components: a clustering system, target-speaker voice activity detection (TS-VAD), and target-speaker ASR (TS-ASR). To enhance the system's performance and applicability, we integrate target-speaker embeddings as prompts within the TS-VAD and TS-ASR modules. Leveraging pre-trained models during system initialization further empowers our approach, granting it superior generalizability and precision. In experiments conducted on the M2MeT2.0 dataset, our integrated PP-MeT system achieves a concatenated minimum permutation character error rate (cp-CER) of only 11.27\% on the test set, achieving the top position in both fixed and open training conditions. We also release our inference system with pre-trained models at website\footnote{https://github.com/XimalayaEverestIntelligentLab/M2MET2.0}.%, underscoring its remarkable performance and effectiveness.
%construct a cascaded system meticulously crafted to address the intricacies of multi-party meetings in dinner party scenario. To begin with, we use a spectral clustering system with pre-trained speaker embedding model to obtain a relatively accurate speaker diarization result. Then, we extract personalized prompt for each speaker, and feed it into our TS-VAD system, to yield a more accurate diarization result for overlapped speech. Finally, we feed personalized prompt and enhanced speech into our TS-ASR system, and generates transcription with speaker labels

%Contribution: our proposed system is as follows. 

The rest of this paper is organized as follows. In Section 2, we detail the architecture of the PP-MeT system. Datasets and experimental setup are described in Section 3. Section 4 presents the experimental results of M2MeT2.0 Challenge test set and our ablation study. Finally, we conclude in Section 5.
%gives the experimental result on M2MeT2.0 testset and our ablation study. Finally, section 5 concludes the paper and presents the future work.
\section{Proposed System Description}
\label{ssec:method}

\begin{figure}[th] 
\centering
\centerline{\includegraphics[width=0.95\linewidth]{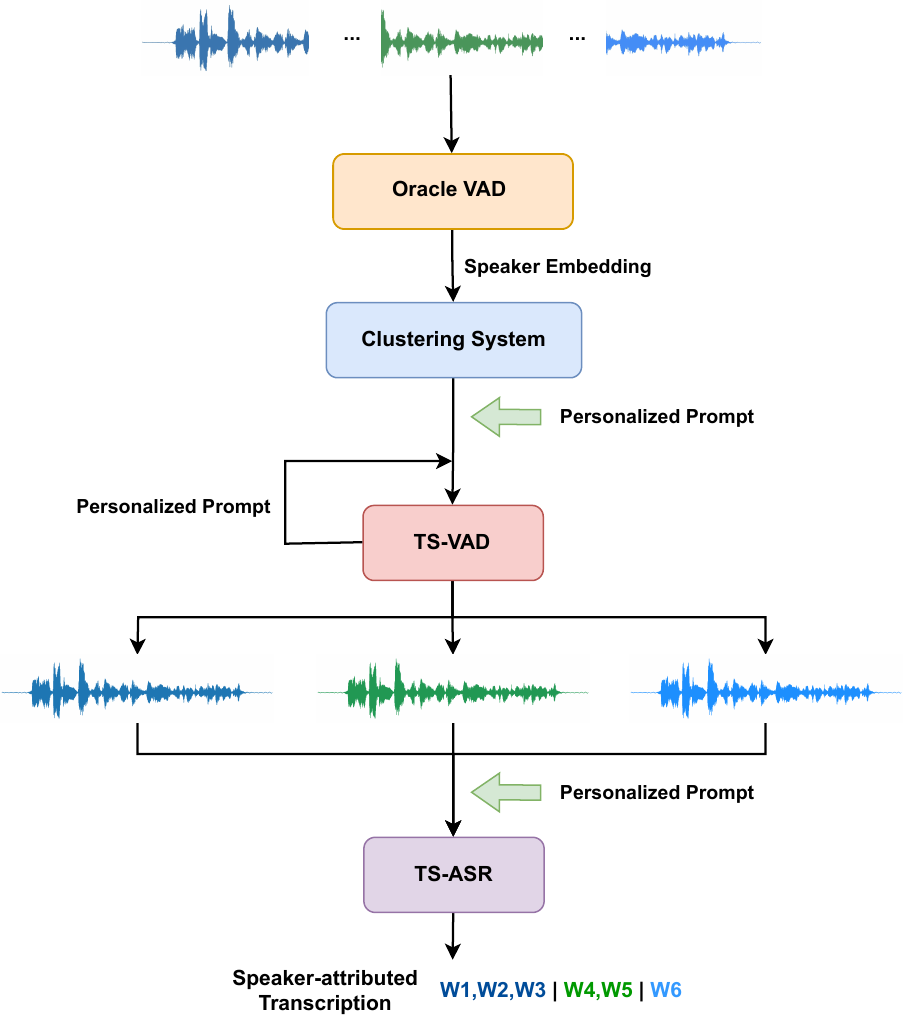}}
\caption{The overview of our proposed PP-MeT system.}
\label{fig:overview}
\vspace{-1em}%
\end{figure}

The overview of our proposed PP-MeT system for the M2MeT2.0 Challenge is shown in Figure~\ref{fig:overview}.

%Our inference system with pretrained models is released at \footnote{https://github.com/aluminumbox/M2MeT2.0}, while the training system is not available yet due to its complexity.
%\cite{XimalayaEverestIntelligentLab}. 
\subsection{Speaker Embedding System}

As M2MeT2.0 encourage the participants to use pre-trained models, we use two pre-trained models\footnote{https://github.com/wenet-e2e/wespeaker/blob/master/docs/pretrained.md} from Wespeaker toolkit~\cite{wang2023wespeaker,he2016deep}. 
One is Resnet34 from the CN-Celeb example, and another is ResNet34-LM, which is obtained by further training ResNet34 with a large-margin technique. We also train a ResNet34 model with Speechbrain toolkit\footnote{https://github.com/speechbrain/speechbrain/tree/develop} to introduce diversity to our speaker embedding model. We will refer to these three speaker embedding models as SV-1/2/3 and the corresponding personalized prompt as Prompt-1/2/3 for simplicity.

\subsection{Clustering System}

Before proceeding to TS-VAD and TS-ASR systems, we need to estimate the number of speakers and initialize personalized prompts using clustering algorithm.

First, we extract voice speech segments based on VAD results for each session. Then we split each segment into subsegments using a fixed 3s window size and 1.5s window-shift. After that, we use speaker embedding model to extract embedding for each subsegment. Finally, we feed the L2-normalized embedding into the clustering algorithm and obtain the speaker number for each session, as well as the label for each subsegment.

We use DOVER-lap toolkit~\cite{raj2021dover}\footnote{https://github.com/desh2608/dover-lap} to merge clustering results from different channel and speaker embedding models. We compare Auto-tuning Spectral Clustering with Normalized Eigen Gap~\cite{park2019auto}(NME-SC) with Agglomerative Hierarchical Clustering(AHC) algorithm. As NME-SC outperforms AHC by a large margin, we use NME-SC algorithm results to initialize personalized prompts. 

After obtaining the clustering system result, for each speaker, we extract speech that contains only the targeted speaker as personalized speech. Then we repeat the speaker embedding extraction steps over the personalized speech and use the mean-pooled L2-normalized speaker embedding as the personalized prompt.

\subsection{TS-VAD System}

As the clustering system can not handle overlap speech, it results in a high miss error in multi-party meeting scenarios. To further reduce DER, we use TS-VAD system to give a more accurate estimate of speaker labels.

\begin{figure}[] 
\centering
\centerline{\includegraphics[width=0.95\linewidth]{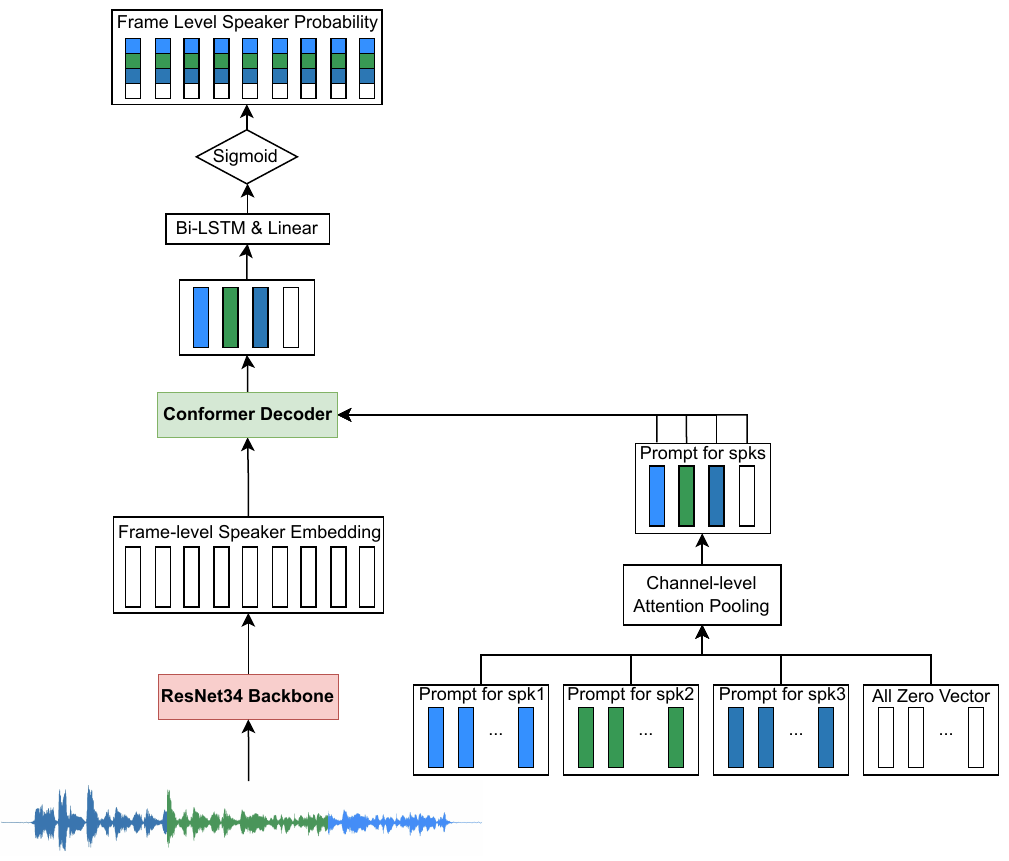}}
\caption{TS-VAD model structure}
\label{fig:TS-VAD}
\vspace{-1em}%
\end{figure}

We use ResNet34 model as backbone for our TS-VAD system, which is the same as that of speaker embedding model. First, we extract the pooling layer input as frame-level speaker embedding. Then we do a stats-pooling with 3-second stride to extract the frame level mean and std feature, and concatenate it with the original frame-level speaker embedding. We do mean-pooling and attention-pooling for frame-level speaker embedding and personalized prompts, respectively. After that, we use a conformer decoder layer to explore the relationship between the frame-level speaker embedding and personalized prompt. We feed the frame-level speaker embedding feature as conformer decoder input, and each personalized prompt as decoder memory. Finally, we concatenate the conformer decoder layer output and use a BiLSTM layer to explore the relationship among each speaker. The BiLSTM output is fed into a fully-connect(FC) layer with a sigmoid activation function to generate the final TS-VAD probability\cite{wang2022cross,medennikov2020target}. The detailed TS-VAD model structure is shown in Figure~\ref{fig:TS-VAD}.

\subsection{TS-ASR system}

Far-Field ASR poses a greater challenge compared to ASR of speech captured by a close-proximity microphone due to the degraded quality of the signal. To address this, we endeavor to engage in speech enhancement. In our practical pursuit, there exist two pivotal components. Firstly, we employ a sophisticated dereverberation method based on weighted prediction error (WPE)~\cite{nakatani2010speech}  to mitigate the effects of late reverberation. In the challenge, we utilize an accelerated GPU-version of WPE, incorporating the following parameters: taps=12, delay=2, iterations=3. Secondly, in order to further attenuate late reverberation and minimize noise interference, the weighted delay-and-sum acoustic beamforming (BeamformIt) method~\cite{anguera2007acoustic} is employed.

\begin{figure}[th]
\centering
\centerline{\includegraphics[width=0.95\linewidth]{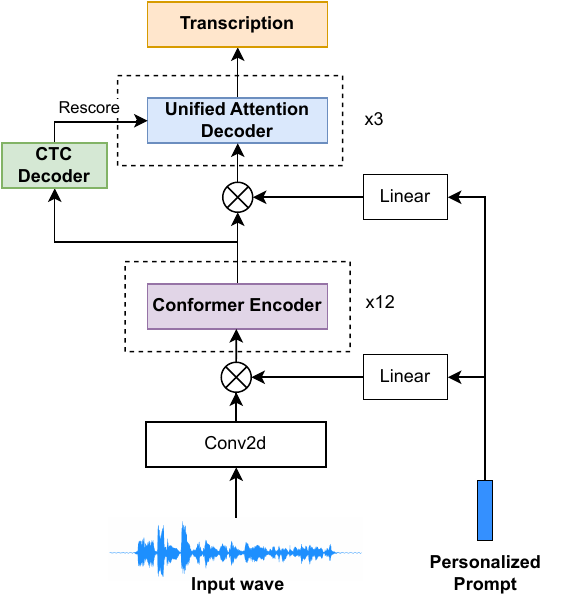}}
\caption{TS-ASR model structure}
\label{fig:TS-ASR}
\vspace{-1em}%
\end{figure}

As M2MeT2.0 requires participants to give transcription for each speaker, we upgrade the traditional ASR model into TS-ASR system with personalized prompt module, which enables it to yield different transcription given different personalized prompt~\cite{moriya2022streaming,sato2021should}. We feed the personalized prompt into a FC layer, and do Hadamard product with the output from the first layer of both asr encoder and decoder. As our TS-ASR model makes little modification to the traditional ASR model, we can easily adapt a pre-trained ASR model into a TS-ASR model. We use Unfied-Conformer~\cite{gulati2020conformer} model pretrained on wenetspeech\footnote{https://github.com/wenet-e2e/wenet/blob/main/docs/pretrained\_models.md} from~\cite{zhang2022wenet} as the TS-ASR model backbone.
The detailed TS-ASR model structure is shown in Figure~\ref{fig:TS-ASR}.

\section{Experimental Setup}
\label{ssec:setup}

\subsection{Datasets}
The original M2MeT1.0 dataset~\cite{yu2022m2met} contains 118.75 hours of speech data in total. The dataset is divided into 104.75 hours for training, 4 hours for development (denoted as Dev 1.0), and 10 hours as test set (denoted as Test 1.0) for scoring and ranking in M2MeT1.0 Challenge. Test 1.0 is used as development set in M2MeT2.0 Challenge. M2MeT2.0 uses a new 10 hours dataset (denote as Test 2.0) as test set.
AISHELL4~\cite{fu2021aishell} is a real-recorded Mandarin speech dataset collected by 8-channel circular microphone array for speech processing in a conference scenario. This dataset consists of 211 recorded meeting sessions, each containing 4 to 8 speakers, with a total length of 120 hours, aiming to bridge the advanced research on multi-speaker processing and the practical application scenarios. 
CN-Celeb~\cite{fan2020cn} is a large-scale speaker recognition dataset collected ‘in the wild’. This dataset contains more than 130, 000 utterances from 1, 000 Chinese celebrities, and covers 11 different genres in the real world.

Both M2MeT and AISHELL4 datasets are far-field multi-channel datasets, while the CN-Celeb dataset is a near-field dataset. Figure~\ref{fig:datapre} shows the data preparation. By Oracle VAD, the non-overlap speech of each speaker is obtained from both near-field and far-field data. Then the personalized prompt is extracted. 
The M2MeT dataset is processed according to the given prior information into continuous voice speech. All far-field multi-channel datasets are pre-processed to generate enhanced data by WPE and BF.
The original far-field 8-channel data and the enhanced data compose the speech of  each speaker, which is used in the next training process.
\iffalse
\begin{figure*}
    \centering
    {\includegraphics[width=17.8cm]{datapre.pdf}}
    \caption{Data preparation before training.}
    \label{fig:datapre}
\end{figure*}
\fi

\begin{figure}
    \centering
    {\includegraphics[width=8.5cm]{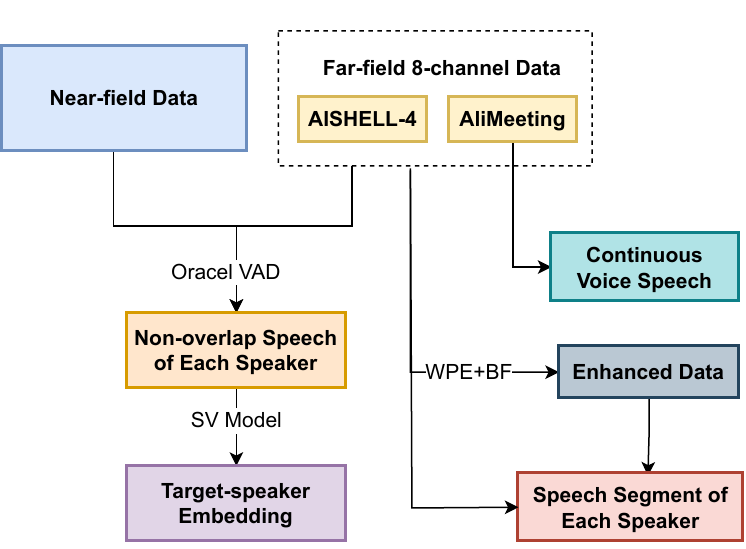}}
    \caption{Data preparation before training.}
    \label{fig:datapre}
\end{figure}

\begin{figure}
    \centering
    {\includegraphics[width=8.5cm]{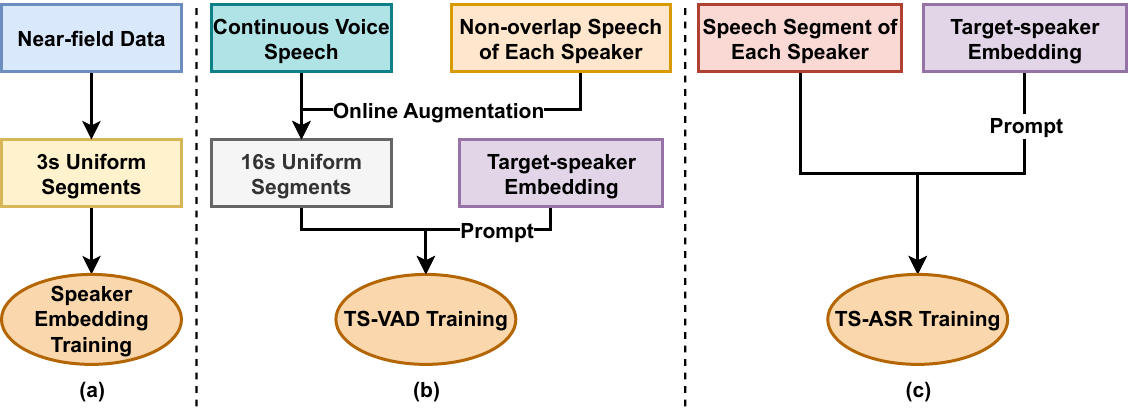}}
    \caption{Data flow in each training process.}
    \label{fig:dataflow}
\end{figure}

The data flow of each training process is shown in Figure~\ref{fig:dataflow}. The near-field data is processed into 3-second uniform segments and used in speaker embedding training. In TS-VAD model training, the continuous voice speech and non-overlap speech with online augmentation are processed into 16-second uniform segments, and the target-speaker embedding is used as a prompt. Moreover, the speech segment of each speaker and the personalized prompt are used in TS-ASR model training.
%speaker model？trainin？

\subsection{System Setup}

For all systems, we use 80-dimension log-mel filter bank (Fbank) feature as input. The Fbank feature is extracted using a 25ms window length and 10ms window shift.

\subsubsection{Speaker embedding system}
we use CN-Celeb data~\cite{fan2020cn} to train our speaker embedding model and split each utterance into 3s uniform length segments. When iterating over all segments, we introduce diversity by randomly offsetting the start frame of the segments from -1.5s to 1.5s. All these three speaker embedding models are trained using AAM softmax loss~\cite{deng2019arcface} and generate 256 dimension speaker embedding as output. We use a cyclical learning rate policy to dynamically adjust the lr for 16 epochs.

\subsubsection{TS-VAD system} 
we use M2MeT2.0 training data and Aishell-4 data for training. For each session, first, we extract and combine all voiced speech as our real training data. Then, for each speaker, we extract and combine speech that contains only the target speaker as personalized speech. Finally, we initialize Prompt-1/2/3 using personalized speech. If the number of speaker is less than 4, we pad Prompt-1/2/3 using zero vectors.

During training, we split the real training data into 16s segments and iterate over each segment. We also do an online data simulation by choosing personalized speech from random speakers to fill up the voiced region of real data\cite{wang2021dku}. It is important that the randomly chosen speakers are from the same session, in case the model learns background noise feature of each session, rather than the essential difference of each speaker.

We train three TS-VAD models based on SV-1/2/3 and Prompt-1/2/3. For all TS-VAD models, we use 2 layers, 256-dimension input, 512-dimension hidden dimension, and 8 heads for the conformer decoder setup. We use 2 layers, 1024 dimension input, and 512 hidden dimensions for BiLSTM setup.

TS-VAD training consists three key stages. In stage 1, we copy the pre-trained speaker embedding parameter into the TS-VAD model, freeze the backbone part and train the model using real and simulated data until convergence with 1e-3 lr. In stage 2, we train the whole model using real and simulated data until convergence with 1e-4 lr. In stage 3, we finetune the whole model only using real data with 1e-5 lr. We choose the model with the lowest DER on Test 1.0 for decoding.

During TS-VAD decoding, we initialize Prompt-1/2/3 from clustering system. We can iterate over the TS-VAD system by re-initialize Prompt-1/2/3 using TS-VAD system outputs.

\subsubsection{TS-ASR system}
We use the WeNet toolkit and its pre-trained Unified-Conformer model on WeNetSpeech as backbone. Since M2MeT2.0 and Aishell-4 training data comprise multiple channels, on one hand, we directly feed the model with raw mean-pooled data, on the other hand, we feed the model with enhanced single-channel data. Additionally, we incorporate speed augmentation techniques during the training process. It is imperative to note that when the audio speed is altered, the corresponding personalized prompt for that particular speed variation should be rendered.

We also train three TS-ASR systems based on Prompt-1/2/3. For all TS-ASR models, we use 12-layer conformer encoder with 512 dimension output, 2048 dimension linear units, and 8 attention heads. We use a 3-layer bi-transformer decoder with 2048 dimension linear units and 8 attention heads. For the personalized prompt module, we feed the 256 dimension personalized prompt into a FC layer, project it into a 512 dimension vector and do a Hadamard product with the first layer output of both encoder and decoder.

TS-ASR training also consists three key stages. In stage 1, we freeze the Unified-Conformer backbone, and only train the personalized prompt module using raw data and enhanced data. In stage 2, we train the whole model with 1e-4 lr. In stage 3, we finetune the whole model with 1e-5 lr using enhanced data.

\section{Experimental Results}
\subsection{Results on M2MeT2.0 Challenge}
M2MeT2.0 challenge uses concatenated minimum permutation character error rate (cp-CER) as the evaluation metric. It computes the minimum CER given all speaker permutations, which requires the system to give the correct transcription and speaker label. The calculation of cp-CER is divided into three steps. First, recognition results and reference transcriptions belonging to the same speaker are concatenated on the timeline in a session. Second, the character error rate (CER) of all permutations of speakers is calculated. Finally, the lowest CER is selected as the cp-CER.

Table~\ref{table:1} presents the cp-CER results of the official baseline and each competition system. Our system achieves 15.05\%, 16.84\%, and 11.27\% cp-CER on Dev 1.0, Test 1.0, and Test 2.0, respectively. Notice that cp-CER on Dev 1.0 and Test 1.0 is achieved using oracle diarization result. We can observe that our PP-MeT model gives better results over the official baseline and achieve up to 30.28\% absolute cp-CER improvement due to the enhanced dataset and advanced model architectures. achieving first place in the challenge.

\begin{table}[h!]
\centering
\begin{tabular}{cccc}% 其中，tabular是表格内容的环境；c表示centering，即文本格式居中；c的个数代表列的个数
\toprule %[2pt]设置线宽     
System   &  Dev 1.0 & Test 1.0 & Test 2.0 \\ %换行
\midrule %[2pt]  
\textbf{PP-MeT (Rank 1st)} & \textbf{15.05} & \textbf{16.84} & \textbf{11.27}  \\
Rank 2nd Team & -- & -- & 18.64  \\
Rank 3rd Team & -- & -- & 22.83 \\
Rank 4th Team & -- & -- & 23.51 \\
Rank 5th Team & -- & -- & 24.82 \\
Official Baseline & 47.4 & 52.57 & 41.55  \\
\bottomrule %[2pt]     
\end{tabular}
\caption{The cp-CER (\%)$\downarrow$ results of each competition system on the M2MeT Dev 1.0, Test 1.0, and Test 2.0.}
\label{table:1}
\end{table}

%Comparing M1 and M2, the data simulation gives a noticeable gain.
\subsection{Ablation Study}

We conduct a detailed ablation study to better understand the contribution of cp-CER from each system, and the significance of pre-trained models.

\begin{table*}[]
%\vspace{-5pt}
\label{tab:der}
\renewcommand{\tabcolsep}{0.055cm}
\renewcommand\arraystretch{1.3}
\begin{tabular}{c|c|cccccccc|c|c}
\bottomrule
Clustering Method    & SV Model & \multicolumn{8}{c|}{Channel 1-8 DER(\%) $\downarrow$}                     & DER (\%)\_Channel $\downarrow$& DER (\%)\_Model $\downarrow$\\ \hline
\multirow{3}{*}{SC}  & SV-1     & 16.87 & 16.21 & 16.89 & 17.49 & 18.61 & 18.00 & \textbf{16.09} & 18.85 & 16.40                         & \multirow{3}{*}{\textbf{15.22}}      \\
                     & SV-2     & \textbf{15.96} & 16.22 & 16.41 & 16.39 & 16.86 & 16.57 & 16.55 & 17.99 & \textbf{15.22}                         &                             \\
                     & SV-3     & 17.26 & \textbf{16.18} & 17.16 & 16.97 & 17.07 & 16.55 & 16.52 & 17.11 & 15.75                         &                             \\ \hline
\multirow{3}{*}{AHC} & SV-1     & 28.22 & 27.11 & 26.92 & 25.79 & 25.90 & 24.52 & 26.61 & 26.25 & 22.95                         & \multirow{3}{*}{22.43}      \\
                     & SV-2     & 26.48 & 23.85 & 26.13 & 26.70 & 25.28 & 25.61 & 24.58 & 26.65 & 22.43                         &                             \\
                     & SV-3     & 25.90 & 26.74 & 29.33 & 27.19 & 27.51 & 27.02 & 26.09 & 25.87 & 22.94                         &                             \\ \bottomrule
\end{tabular}
\caption{DER Results for each clustering system on Test 1.0}
\label{table:2}
\end{table*}
\subsubsection{Clustering System}

As clustering system gives the estimate of speaker number and rough speaker label, its performance determines the superior limit of the whole PP-MeT system. In Table~\ref{table:2}, we study the impact of different speaker embedding models and clustering algorithms in clustering systems.

SV-1/2/3 achieves 7.13\%, 6.49\%, and 7.06\% on CN-Celeb dev trials, respectively. The threshold for AHC clustering is tuned on Dev 1.0. Results show that given each model and channel, NME-SC outperforms AHC significantly. DOVER-lap makes the clustering result more stable by leveraging clustering results from different channels and models. 

As the accuracy of speaker embedding directly affects the quality of speaker embedding, DER relates to speaker embedding performance evidently. The lowest DER is achieved by SV-2, which also achieves the lowest EER on CN-Celeb trials.

\subsubsection{TS-VAD system}
In Table~\ref{table:3}, we study the impact of pre-trained speaker embedding model and different model architectures in TS-VAD system. 

Results show that pre-trained model contributed heavily to the performance of TS-VAD system. If TS-VAD model backbone parameter is randomly initialized, it only achieves 13.28\% DER on Test 1.0, which is only slightly better than that of clustering system. Also, TS-VAD model backbone should match that of the personalized prompt. If we initialize the TS-VAD model backbone parameter using pre-trained ECAPA-TDNN speaker embedding model and train with Prompt-1. It achieves 7.68\% DER, which is much worse than its counterpart using matched speaker embedding model and prompt. The above results demonstrate the importance of pre-trained models in TS-VAD system, and using matched speaker embedding model for initialization and personalized prompt makes it easier to explore the relationship between frame-level speaker embedding and personalized prompt.

We can also observe that the DER drops moderately if we iterate the TS-VAD system by refining Prompt-1/2/3 using TS-VAD system output. 

\begin{table*}[]
\renewcommand{\tabcolsep}{0.02cm}
\renewcommand\arraystretch{1.3}
\begin{tabular}{c|c|c|ccccccc}
\bottomrule
 Initialization Model Parameter & Personalized Prompt & DER (\%)\_Iter0 $\downarrow$ & \multicolumn{1}{c|}{DER (\%)\_Model $\downarrow$}       & \multicolumn{1}{c|}{DER (\%)\_Iter1 $\downarrow$} & DER (\%)\_Model $\downarrow$       &  &  &  &  \\ \hline
 SV-1                   & Prompt-1    & 5.22            & \multicolumn{1}{c|}{\multirow{3}{*}{\textbf{3.19}}} & \multicolumn{1}{c|}{4.87}            & \multirow{3}{*}{\textbf{2.99}} &  &  &  &  \\ \cline{1-1}
SV-2                   & Prompt-2    & 4.52            & \multicolumn{1}{c|}{}                      & \multicolumn{1}{c|}{4.25}            &                       &  &  &  &  \\ \cline{1-1}
 SV-3                   & Prompt-3    & \textbf{4.02}            & \multicolumn{1}{c|}{--}                      & \multicolumn{1}{c|}{\textbf{3.64}}            &                       &  &  &  &  \\ \cline{1-1} \cline{4-7}
 Random                  & Prompt-1    & 13.28           & \multicolumn{3}{c}{--}                                                                                    &  &  &  &  \\ \hline
 ECAPA                   & Prompt-1    & 7.68            & \multicolumn{3}{c}{--}                                                                                    &  &  &  &  \\ \bottomrule
\end{tabular}
\caption{DER Results for each TS-VAD model on Test 1.0}
\label{table:3}
\end{table*}

\iffalse
\begin{table}[h!]
\centering
\begin{tabular}{ccc}% 其中，tabular是表格内容的环境；c表示centering，即文本格式居中；c的个数代表列的个数
\toprule %[2pt]设置线宽     
TS-VAD System  & DER \\ %换行
\midrule %[2pt]  
ResNet34 backbone & 13.28 \\
\bottomrule %[2pt]     
\end{tabular}
\caption{DER Results for TS-VAD on testset without pretrained parameters}
\label{table:3}
\end{table}

\begin{table}[h!]
\centering
\begin{tabular}{ccc}% 其中，tabular是表格内容的环境；c表示centering，即文本格式居中；c的个数代表列的个数
\toprule %[2pt]设置线宽     
TS-VAD System  & DER \\ %换行
\midrule %[2pt]  
ResNet34 backbone & 7.68 \\
\bottomrule %[2pt]     
\end{tabular}
\caption{DER Results for TS-VAD on testset with unfitted model backbone and personalized speaker embedding}
\label{table:3}
\end{table}

\begin{table}[h!]
\centering
\begin{tabular}{ccc}% 其中，tabular是表格内容的环境；c表示centering，即文本格式居中；c的个数代表列的个数
\toprule %[2pt]设置线宽     
TS-VAD System  & DER & DER(iter1) \\ %换行
\midrule %[2pt]  
Wespeaker ResNet34 backbone & 5.22 & 4.87 \\
Wespeaker ResNet34-LM backbone & 4.52 & 4.25 \\
Speechbrain ResNet34 backbone & 4.02 & 3.64 \\
Dover lap & 3.19 & 2.99 \\
\bottomrule %[2pt]     
\end{tabular}
\caption{DER Results for TS-VAD on testset}
\label{table:3}
\end{table}
\fi

\subsubsection{TS-ASR System}
In Table~\ref{table:4}, we study the impact of pre-trained models and personalized prompts in TS-ASR system. 

First, we try to finetune the pre-trained unified-conformer ASR model directly without any structure modification. Results show that the pre-trained ASR model achieves 32.63\% and 35.89\% cp-CER on Dev 1.0 and Test 1.0. After finetuning the model on M2MET2.0 and Aishell-4 data, the cp-CER drops to 22.55\% and 26.43\%, respectively. However, the cp-CER improvement is largely due to the model performance on nonoverlap speech. It fails to decrease further because the traditional ASR model cannot handle overlap speech.

Then, we try to train the TS-ASR model from scratch with Prompt-1. However, the TS-ASR model with a unified-conformer backbone fails to converge. This demonstrates the necessity of pre-trained ASR model backbone in our TS-ASR system.

Finally, we train three TS-ASR models based on the Prompt-1/2/3. cp-CER on TS-ASR model with pretrained ASR model backbone and Prompt-1/2/3 drops dramatically both on Dev 1.0 and Test 1.0. The result shows that pre-trained ASR model with Prompt-2 achieves the lowest cp-CER, which means that the performance of pre-trained speaker embedding model also affects the performance of TS-ASR on overlapped speeches.

We also try to finetune the TS-ASR model further using LF-MMI with k2 toolkit\footnote{https://github.com/k2-fsa/k2}, and introducing LM information by decoding with HLG. However, the cp-CER fails to drop on both Dev 1.0 and Test 1.0. This is due to the fact that in multi-party meeting scenario, the transcription from each session is highly irrelevant. External LM information can not help to decrease cp-CER.

In Table~\ref{table:4}, \textit{Test 1.0} cp-CER is calculated using segments and prompts from TS-VAD system. The gap between cp-CER of \textit{Test 1.0} and Test 1.0 means the degradation introduced by TS-VAD system, which is approximately 2\%. We obtain final results by leveraging each system results using SCTK rover toolkit\footnote{https://github.com/usnistgov/SCTK}.

\begin{table}[h!]
\centering
\begin{tabular}{cccc}% 其中，tabular是表格内容的环境；c表示centering，即文本格式居中；c的个数代表列的个数
\toprule %[2pt]设置线宽     
 Personalized Prompt  & Dev 1.0 & Test 1.0  & \textit{Test 1.0} \\ %换行
\midrule %[2pt]  
%unified-transformer & -- & wespeaker ResNet34-LM & 26.13 & 28.56 & -- \\
%%%%%%不用pre-train模型初始化的话，在unified transformer或者conformer上要么不收敛，要么结果很差，
%unified-conformer & -- & wespeaker ResNet34-LM & -- & -- & -- \\
 -- & 22.55 & 26.43 & -- \\

 Prompt-1 & 15.35 & 17.20 & 19.45 \\
 Prompt-2 & \textbf{15.13} & \textbf{17.08} & \textbf{19.06} \\
 Prompt-3 & 15.28 & 17.16 & 19.06 \\
 Rover & \textbf{15.05} & \textbf{16.84} & \textbf{18.92} \\
\bottomrule %[2pt]     
\end{tabular}
\caption{cp-CER (\%) results for each TS-ASR model on Dev 1.0, Test 1.0 and \textit{Test 1.0}.}
\label{table:4}
\end{table}

\section{Conclusion}
\label{sec:conclu}

%In this paper, we describe our PP-MET system in the Multi-channel Multi-party Meeting Transcription Challenge 2.0 (M2MeT2.0), which aims to solve the automatic speech recognition(ASR) task in a multi-party meeting scenario.
%Specifically, our proposed integrated PP-MET system consists of a clustering system, TS-VAD, and TS-ASR system. Moreover, we utilize personalized speaker embedding as a prompt in TS-VAD and TS-ASR in our proposed integrated meeting transcription system.
%To improve the generalizability of our meeting transcription system, pre-trained model is used for initialization, which is also a fast adaptation among all the modules.

%In the future, we will explore the possibility of expanding personalized prompts on time axis. Also, we will improve the TS-ASR model by jointly training speaker embedding backbone with ASR system.

In this paper, we present our PP-MET system for the Multi-channel Multi-party Meeting Transcription Challenge 2.0 (M2MeT2.0) to address the ASR task in a multi-party meeting scenario. Compared with the other conventional systems, we incorporate target-speaker embedding as a personalized prompt in both TS-VAD and TS-ASR stage. Moreover, to further enhance the system's robustness and reduce the training cost, pre-trained models are used in our system's initialization, enabling fast adaptation across all modules. Experimental results shows proposed system outperforms conventional systems by a large margin.

In future work, we plan to explore the potential of expanding personalized prompts on the time axis. Additionally, we aim to enhance the TS-ASR model by jointly training the speaker embedding module with the ASR backbone, further improving its performance.

\bibliographystyle{IEEEbib}
\bibliography{mybib}

\end{document}